\documentclass[a4paper]{jpconf}
\usepackage{graphicx}
\usepackage{amsmath,amssymb}
\newcommand{\bra}[1]{\langle {#1} |}
\newcommand{\ket}[1]{| {#1} \rangle}
\newcommand{\inproduct}[2]{\langle #1 | #2 \rangle}
\newcommand{\ainproduct}[2]{\langle\langle #1 | #2 \rangle\rangle}

\begin{document}
\title{Linear response calculation using the canonical-basis
TDHFB with a schematic pairing functional}

\author{Shuichiro Ebata$^{1,2}$,
Takashi Nakatsukasa$^{1,3}$,
and Kazuhiro Yabana$^{1,2,3}$}

\address{$^1$ RIKEN Nishina Center, Wako-shi 351-0198, Japan}
\address{$^2$ Graduate School of Pure and Applied Sciences,
University of Tsukuba, Tsukuba 305-8571, Japan}
\address{$^3$ Center for Computational Sciences, University of Tsukuba,
Tsukuba 305-8571, Japan}


\begin{abstract}
A canonical-basis formulation
of the time-dependent Hartree-Fock-Bogoliubov (TDHFB) theory
is obtained with an approximation
that the pair potential is assumed to be diagonal in the
time-dependent canonical basis.
The canonical-basis formulation significantly reduces the computational cost.
We apply the method to linear-response calculations for even-even nuclei.
$E1$ strength distributions for proton-rich Mg isotopes
are systematically calculated.
The calculation suggests strong Landau damping of giant dipole resonance
for drip-line nuclei.
\end{abstract}

\section{Introduction}
The time-dependent Hartree-Fock (TDHF) theory has been extensively utilized to
study nuclear many-body dynamics \cite{Neg82}.
Recently, it has been revisited with modern energy density functionals
and more accurate description of nuclear properties
has been achieved \cite{NY05,Mar05,UO06,UO07,WL08,CC09}.
The TDHF theory uses only occupied
orbitals, number of which is equal to the number of particles ($N$).
However, it neglects the residual interactions in
particle-particle and hole-hole channels,
which are important for properties of open-shell heavy nuclei.
It is well-known that the time-dependent Hartree-Fock-Bogoliubov (TDHFB)
theory \cite{BR86} properly takes into account the pairing correlations.
The TDHFB equation is formulated in a similar manner to the TDHF,
however it requires us to calculate the time evolution of quasi-particle
orbitals, number of which is, in principle, infinite.
Therefore, the practical calculations with the TDHFB are very limited
\cite{HN07,ASC08}.

Very recently, we have proposed a possible approximation for the TDHFB theory
\cite{Eba10}.
This is a time-dependent version of the BCS approximation \cite{RS80} for
the Hartree-Fock-Bogoliubov theory.
Namely, we neglect off-diagonal elements of the pair potential in the
time-dependent canonical basis.
We show that this approximation results in significant reduction of
the computational task.
We call the equations obtained with this approximation,
``Canonical-basis TDHFB'' (Cb-TDHFB) equations.
We apply the method to the linear-response calculations using
the full Skyrme functionals of the parameter set of SkM*,
and discuss properties of the $E1$ strength distribution
in even-even Mg isotopes.

The paper is organized as follows.
In Sec.~\ref{sec:formalism}, we present the basic equations of the present
method and their derivation.
It is emphasized that the basic equations possess a gauge invariance.
A schematic choice of the pairing functional leads to
violation of the gauge invariance, which requires us to choose
a specific gauge to minimize the violation.
In Sec.~\ref{sec:numerical_results},
we present numerical results of the real-time calculations of the
linear response for Mg isotopes.
Finally, the conclusion is given in Sec.~\ref{sec:summary}.

\section{Formalism of the Cb-TDHFB theory}
\label{sec:formalism}

In this section, we show the basic equations of the Cb-TDHFB and
recapitulate their derivation.
See Ref. \cite{Eba10} for more details.

\subsection{Basic equations}

Let us first show the Cb-TDHFB equations we derive in the followings.
\begin{subequations}
\label{Cb-TDHFB}
\begin{eqnarray}
\label{dphi_dt}
&&i\frac{\partial}{\partial t} \ket{\phi_k(t)} =
(h(t)-\eta_k(t))\ket{\phi_k(t)} , \quad\quad
i\frac{\partial}{\partial t} \ket{\phi_{\bar k}(t)} =
(h(t)-\eta_{\bar k}(t))\ket{\phi_{\bar k}(t)} , \\
\label{drho_dt}
&&
i\frac{d}{dt}\rho_k(t) =
\kappa_k(t) \Delta_k^*(t)
-\kappa_k^*(t) \Delta_k(t) , \\
\label{dkappa_dt}
&&
i\frac{d}{dt}\kappa_k(t) =
\left(
\eta_k(t)+\eta_{\bar k}(t)
\right) \kappa_k(t) +
\Delta_k(t) \left( 2\rho_k(t) -1 \right) .
\end{eqnarray}
\end{subequations}
These basic equations determine the time evolution of
the canonical states, $\ket{\phi_k(t)}$ and $\ket{\phi_{\bar k}(t)}$,
their occupation, $\rho_k(t)$, and pair probabilities, $\kappa_k(t)$.
The real functions of time, $\eta_k(t)$ and $\eta_{\bar k}(t)$,
are arbitrary and associated with the gauge degrees of freedom.
The time-dependent pairing gaps,
$\Delta_k(t)$, which are given in Eq. (\ref{Delta_k}),
are similar to the BCS pairing gap \cite{RS80} except for
the fact that the canonical pair of states are no longer
related to each other by time reversal.
Although we use the same symbols, $(\rho, \kappa, \Delta)$,
for matrixes in Eqs. (\ref{TDHFB_1}) and (\ref{TDHFB_2}),
the quantities in the Cb-TDHFB equations are only their diagonal elements
with a single index for the canonical states $k$.
It should be noted that
similar equations can be found
in Ref.~\cite{BF76} for a simple pairing energy functional.

\subsection{Derivation of the basic equations}

We now derive the Cb-TDHFB equations starting from the generalized
density-matrix formalism.
The TDHFB equation can be written in terms of the generalized density
matrix $R(t)$ as \cite{BR86}
\begin{equation}
i\frac{\partial}{\partial t} R(t) = \left[ {\cal H}(t), R(t) \right] .
\end{equation}
This is equivalent to the following equations for
one-body density matrix, $\rho(t)$,
and the pairing-tensor matrix, $\kappa(t)$.
\begin{subequations}
\begin{eqnarray}
\label{TDHFB_1}
i\frac{\partial}{\partial t}\rho(t) &=&
  [h(t),\rho(t)] + \kappa(t) \Delta^*(t) - \Delta(t) \kappa^*(t) ,\\
\label{TDHFB_2}
i\frac{\partial}{\partial t}\kappa(t) &=&
 h(t)\kappa(t)+\kappa(t) h^*(t) + \Delta(t) (1-\rho^*(t)) - \rho(t) \Delta(t) .
\end{eqnarray}
\end{subequations}
Here, $h(t)$ and $\Delta(t)$ are single-particle Hamiltonian and
pair potential, respectively.

At each instant of time,
we may diagonalize
the density operator $\hat{\rho}$ in
the orthonormal canonical basis, $\{ \phi_k(t), \phi_{\bar k}(t)\}$
with the occupation probabilities $\rho_k$.
Then, the TDHFB state is expressed in the canonical (BCS) form as
\begin{equation}
\ket{\Psi(t)}=\prod_{k>0} \left\{
u_k(t) + v_k(t) c_k^\dagger(t) c_{\bar k}^\dagger(t) \right\} \ket{0} .
\end{equation}
For the canonical states, we use the alphabetic indexes such as $k$
for half of the total space indicated by $k>0$.
For each state with $k>0$, there exists a ``paired'' state ${\bar k}<0$
which is orthogonal to all the states with $k>0$.
The set of states $\{ \phi_k, \phi_{\bar k}\}$ generate the whole
single-particle space.
We use the Greek letters $\mu,\nu,\cdots$ for indexes of
an adopted representation (complete set) for the single-particle states.
Using the following notations,
\begin{eqnarray}
\ainproduct{\mu\nu}{\phi_k(t)\phi_{\bar k}(t)} &\equiv&
\inproduct{\mu}{\phi_k(t)}\inproduct{\nu}{\phi_{\bar k}(t)}-
\inproduct{\mu}{\phi_{\bar k}(t)}\inproduct{\nu}{\phi_k(t)} , \\
\hat{\pi}_k(t) &\equiv& \ket{\phi_k(t)}\bra{\phi_k(t)}
  + \ket{\phi_{\bar k}(t)}\bra{\phi_{\bar k}(t)} ,
\end{eqnarray}
the density and the pairing-tensor matrixes
are expressed as
\begin{eqnarray}
\label{rho_mn}
\rho_{\mu\nu}(t)&=& \sum_{k>0}
\rho_k(t) \bra{\mu}\hat{\pi}_k(t)\ket{\nu} , \\
\label{kappa_mn}
\kappa_{\mu\nu}(t) &=& \sum_{k>0} \kappa_k(t)
\ainproduct{\mu\nu}{\phi_k(t)\phi_{\bar k}(t)} ,
\end{eqnarray}
where $\rho_k(t)=|v_k(t)|^2$ and $\kappa_k(t)=u_k^*(t) v_k(t)$.
It should be noted that the canonical pair of states,
$\ket{\phi_k(t)}$ and $\ket{\phi_{\bar k}(t)}$,
are assumed to be orthonormal
but not necessarily related with each other by the time reversal,
$\ket{\phi_{\bar k}} \neq T\ket{\phi_k}$.

We can invert Eqs. (\ref{rho_mn}) and
(\ref{kappa_mn}) for $\rho_k$ and $\kappa_k$,
\begin{eqnarray}
\label{rho_k}
\rho_k(t)&=&
 \sum_{\mu\nu}
 \inproduct{\phi_k(t)}{\mu}\rho_{\mu\nu}(t)\inproduct{\nu}{\phi_k(t)}
 =\sum_{\mu\nu}
 \inproduct{\phi_{\bar k}(t)}{\mu}\rho_{\mu\nu}(t)
  \inproduct{\nu}{\phi_{\bar k}(t)} , \\
\label{kappa_k}
\kappa_k(t)&=&
\frac{1}{2} \sum_{\mu\nu}
\ainproduct{\phi_k(t)\phi_{\bar k}(t)}{\mu\nu} \kappa_{\mu\nu}(t) .
\end{eqnarray}
The derivative of $\rho_k(t)$ with respect to time $t$ leads to
\begin{eqnarray}
i\frac{d}{dt}\rho_k(t) &=&\sum_{\mu\nu}
\inproduct{\phi_k(t)}{\mu}i\frac{d\rho_{\mu\nu}}{dt}\inproduct{\nu}{\phi_k(t)}
+i\rho_k(t) \frac{d}{dt} \inproduct{\phi_k(t)}{\phi_k(t)} \nonumber\\
&=&\frac{1}{2}\sum_{\mu\nu} \left\{
\kappa_k(t) 
\Delta_{\mu\nu}^*(t)\ainproduct{\nu\mu}{\phi_k(t)\phi_{\bar k}(t)}
+
 \kappa_k^*(t)
\Delta_{\mu\nu}(t) \ainproduct{\phi_k(t)\phi_{\bar k}(t)}{\mu\nu}
\right\} .
\end{eqnarray}
We used the assumption of norm conservation and
the TDHFB equation (\ref{TDHFB_1}).
This can be rewritten in the simple form of Eq. (\ref{drho_dt})
with the definition of the pairing gap,
\begin{equation}
\label{Delta_k}
\Delta_k(t) \equiv
-\frac{1}{2} \sum_{\mu\nu}
\Delta_{\mu\nu}(t)
 \ainproduct{\phi_k(t)\phi_{\bar k}(t)}{\mu\nu} .
\end{equation}
In the same way, we evaluate the time derivative of $\kappa_k(t)$ as
\begin{equation}
i\frac{d}{dt}\kappa_k(t)=\frac{1}{2}\sum_{\mu\nu}
\ainproduct{\phi_k(t)\phi_{\bar k}(t)}{\mu\nu}
i\frac{d\kappa_{\mu\nu}}{dt}
+i\kappa_k(t) \left( \inproduct{\frac{d\phi_k}{dt}}{\phi_k(t)}
                   +\inproduct{\frac{d\phi_{\bar k}}{dt}}{\phi_{\bar k}(t)}
\right) .
\end{equation}
Then, using the TDHFB equation (\ref{TDHFB_2}), we obtain
Eq. (\ref{dkappa_dt}) with the real gauge functions
\begin{equation}
\eta_k(t) \equiv \bra{\phi_k(t)}h(t)\ket{\phi_k(t)}
+i\inproduct{\frac{\partial\phi_k}{\partial t}}{\phi_k(t)} .
\end{equation}
These functions control time dependence of phase for the canonical states,
which are basically arbitrary.

So far, the derivation is based on the TDHFB equations,
and no approximation beyond the TDHFB is introduced.
However, to obtain simple equations for time evolution of
the canonical basis, we need to 
introduce an assumption (approximation) that
the pair potential is written as
\begin{equation}
\label{Delta_mn}
\Delta_{\mu\nu}(t) = -\sum_{k>0}
\Delta_k(t) \ainproduct{\mu\nu}{\phi_k(t)\phi_{\bar k}(t)} .
\end{equation}
This satisfies Eq. (\ref{Delta_k}),
but in general,
Eq. (\ref{Delta_k}) can not be inverted because
the two-particle states $\ket{\phi_k\phi_{\bar k}}$
do not span the whole space.
In other words, we only take into account the pair potential
of the ``diagonal'' parts in the canonical basis,
$\Delta_{k{\bar l}}=-\Delta_k\delta_{kl}$.
In the stationary limit ($\ket{\phi_{\bar k}}=T\ket{\phi_k}$),
this is equivalent to the ordinary BCS approximation \cite{RS80}.
With the approximation of Eq. (\ref{Delta_mn}),
it is easy to see that
the TDHFB equations, \eqref{TDHFB_1} and \eqref{TDHFB_2},
are consistent with
Eqs. \eqref{Cb-TDHFB}.

\subsection{Properties of the Cb-TDHFB equations}
\label{sec:properties}

The Cb-TDHFB equations, \eqref{Cb-TDHFB}, 
are invariant with respect to the gauge transformation with
arbitrary real functions, $\theta_k(t)$ and $\theta_{\bar k}(t)$.
\begin{eqnarray}
\label{gauge_transf_1}
\ket{\phi_k}\rightarrow e^{i\theta_k(t)}\ket{\phi_k}
\quad &\mbox{and}& \quad
\ket{\phi_{\bar k}}\rightarrow e^{i\theta_{\bar k}(t)}\ket{\phi_{\bar k}}
\\
\label{gauge_transf_2}
\kappa_k\rightarrow e^{-i(\theta_k(t)+\theta_{\bar k}(t))}\kappa_k
\quad &\mbox{and}& \quad
\Delta_k\rightarrow e^{-i(\theta_k(t)+\theta_{\bar k}(t))}\Delta_k
\end{eqnarray}
simultaneously with
$$
\eta_k(t)\rightarrow \eta_k(t)+\frac{d\theta_k}{dt}
\quad \mbox{and} \quad
\eta_{\bar k}(t)\rightarrow \eta_{\bar k}(t)+\frac{d\theta_{\bar k}}{dt} .
$$
The phase relations of Eq. (\ref{gauge_transf_2}) are obtained from
Eqs. (\ref{kappa_k}) and (\ref{Delta_k}).
It is now clear that the arbitrary real functions,
$\eta_k(t)$ and $\eta_{\bar k}(t)$,
control time evolution of the phases of
$\ket{\phi_k(t)}$, $\ket{\phi_{\bar k}(t)}$, $\kappa_k(t)$, and
$\Delta_k(t)$.

In addition to the gauge invariance,
the Cb-TDHFB equations possess the following properties.
\begin{enumerate}
\item Conservation law
 \begin{enumerate}
 \item Conservation of orthonormal property of the canonical states
 \item Conservation of average particle number
 \item Conservation of average total energy
 \end{enumerate}
\item The stationary solution corresponds to the HF+BCS solution.
\item Small-amplitude limit
 \begin{enumerate}
 \item The Nambu-Goldstone modes are zero-energy normal-mode solutions.
 \item If the ground state is in the normal phase, the equations are
         identical to the particle-hole, particle-particle, and hole-hole
         RPA with the BCS approximation.
 \end{enumerate}
\end{enumerate}

\subsection{Energy functionals and numerical procedure}

We adopt a Skyrme functional with the SkM* parameter set for the
particle-hole channels.
For the pairing energy functional, we adopt a simple functional of
a form
\begin{equation}
\label{E_G}
E_g(t)=-\sum_{k,l>0} G_{kl} \kappa_k^*(t) \kappa_l(t) ,
=-\sum_{k>0} \kappa_k^*(t) \Delta_k(t) ,
\quad
\Delta_k(t)= \sum_{l>0} G_{kl} \kappa_l(t) ,
\end{equation}
where $G_{kl}=G f(\epsilon_k^0) f(\epsilon_l^0)$ with $G=0.6$ MeV.
The smooth cut-off function $f(\epsilon_k^0)$,
whose explicit form can be found in Ref.~\cite{Eba10},
depends on the single-particle energy
of the canonical state $k$ at the HF+BCS ground state.
A drawback of the simple functional \eqref{E_G} is that
it violates the gauge invariance,
which results in breakdown of some of nice properties shown in
Sec.~\ref{sec:properties}.
However, it is shown that all these properties can be recovered by
choosing a special gauge condition \cite{Eba10}
\begin{equation}
\label{gauge_fix}
\eta_k(t)=\epsilon_k(t)=\bra{\phi_k(t)} h(t) \ket{\phi_k(t)}, \quad
\eta_{\bar k}(t)=\epsilon_{\bar k}(t) 
=\bra{\phi_{\bar k}(t)} h(t) \ket{\phi_{\bar k}(t)} .
\end{equation}

For numerical calculations, we extended the computer program of the
TDHF in the three-dimensional coordinate-space representation \cite{NY05}
to include the pairing correlations.
The ground state is first constructed by the HF+BCS calculation.
Then, we solve the Cb-TDHFB equations in real time, under
a weak impulse isovector dipole field, yielding a time-dependent E1
moment, $D_{E1}(t)$.
To obtain the $E1$ strength distribution,
we perform the spectral analysis with
an exponential smoothing with $\Gamma=1$ MeV:
$D_{E1}(t) \rightarrow D_{E1}(t)e^{-\Gamma t/2}$.
The readers should refer to Ref. \cite{Eba10} for more details.

\section{Electric dipole strength distribution in proton-rich Mg isotopes}
\label{sec:numerical_results}

In Table \ref{tab: gs_properties}, the ground-state deformations,
pairing gaps, and chemical potentials are listed for stable to proton-rich
Mg isotopes.
In the present calculation with SkM*, $^{18}$Mg turns out to be
bound with a small binding energy of 200 keV.
Thus, we include this nucleus in our calculation as a ``fictitious'' proton
halo nucleus.
The neutron pairing gap is absent for all these nuclei.
The proton gap also vanishes for nuclei with prolate shapes,
$\beta=0.3\sim 0.4$.

\begin{table}[tb]
\caption{Ground-state properties of Mg isotopes calculated with
the SkM* functional;
quadrupole deformation parameters $(\beta,\gamma)$,
pairing gaps for neutrons and protons $(\Delta_n,\Delta_p)$,
chemical potentials for neutrons and protons $(\lambda_n,\lambda_p)$.
In the case of normal phase ($\Delta=0$),
we define the chemical potential as the
single-particle energy of the highest occupied orbital.
The pairing gaps and chemical potentials are given in units of MeV.}
\label{tab: gs_properties}
\begin{center}
\begin{tabular}{c|lcllrr}
\br
  & $\beta$ & $\gamma$ & $\Delta_n$ & $\Delta_p$ & $-\lambda_n$ & $-\lambda_p$
\\ \hline
  $^{18}$Mg & 0.31 & 0$^\circ$  & 0.0  & 0.0  & 25.59 & 0.20 \\
  $^{20}$Mg & 0.0  & $-$        & 0.0  & 1.13 & 20.53 & 2.83 \\
  $^{22}$Mg & 0.38 & 0$^\circ$  & 0.0  & 0.0  & 16.31 & 6.42 \\
  $^{24}$Mg & 0.39 & 0$^\circ$  & 0.0  & 0.0  & 14.12 & 9.51 \\
  $^{26}$Mg & 0.20 & 54$^\circ$ & 0.0  & 0.86 & 13.08 & 11.23 \\
  $^{28}$Mg & 0.0  & $-$        & 0.0  & 1.03 &  9.21 & 13.30 \\
\br
\end{tabular}
\end{center}
\end{table}

\begin{figure}[t]
\centerline{
\includegraphics[width=7.6cm,angle=-90]{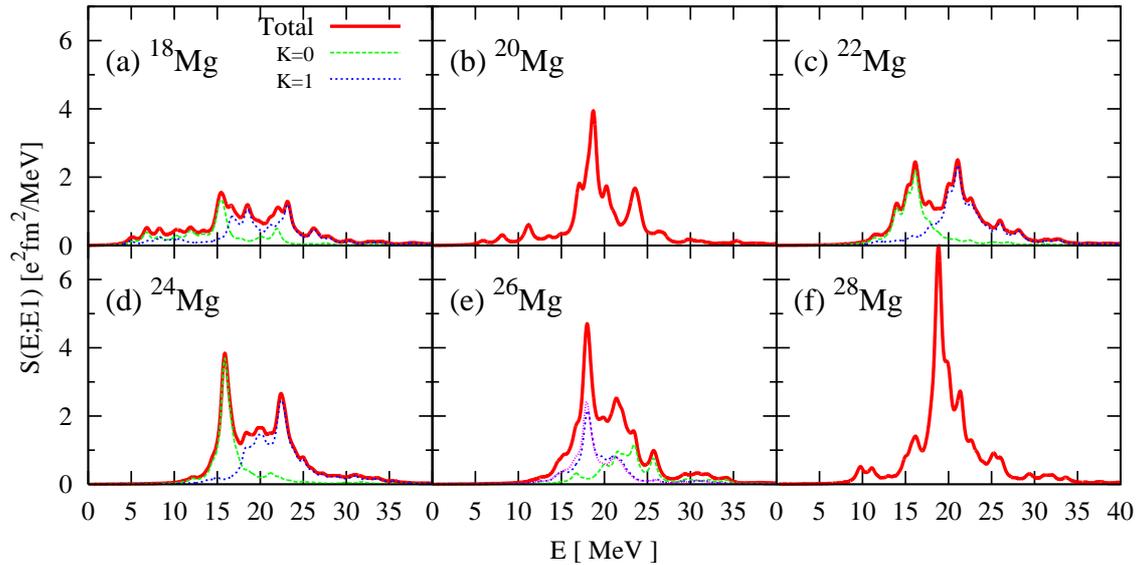}
}
\caption{\label{fig:Mg_E1}
Calculated $E1$ strength distribution for even-even Mg isotopes
($N=8\sim 16$).
For deformed nuclei, the total strength is decomposed
into $K=0$ (green dashed line) and $K=1$
(blue dotted line) components.
The $z$-axis is chosen as the symmetry axis for axially deformed cases.
The smoothing parameter of $\Gamma=1$ MeV is used.
}
\end{figure}

Calculated $E1$ strength distributions are shown in Fig.~\ref{fig:Mg_E1}.
The double-peak structure of the giant dipole resonance (GDR)
due to the deformation splitting
can be seen in $^{22,24}$Mg.
The $K=0$ peak is located around 15 MeV and the $K=1$ is near 22 MeV.
In contrast, for $^{18}$Mg,
although the ground state is deformed in a prolate shape
with $\beta\approx 0.3$, the double-peak structure is not clearly seen.
In this nucleus, the $E1$ strength of both the $K=0$ and $K=1$ components
are fragmented into a wide range of energy.
Previously, we calculated the $E1$ strength distribution in neutron-rich
Mg isotopes \cite{Eba10}, and found the similar damping effects near the
drip line.
This strong Landau damping near the drip lines may be understood by the
high level density of one-particle-one-hole (1p1h) states near the GDR energy.

The low-energy $E1$ strength in $E<10$ MeV is negligible for
stable nuclei ($^{24,26}$Mg).
For the neutron-rich side, we see a small low-energy peak in $^{28}$Mg.
Since the neutron separation energy is still sizable (about 9 MeV),
we assume that this is due to the occupation of the neutron $s_{1/2}$ orbital
which is spatially extended.
For the proton-rich (neutron-deficient) side, the low-energy strength
is seen in $^{18,20}$Mg.
This should be due to the weak binding of the last-occupied proton $d_{5/2}$
orbitals, since
the calculated proton separation energies are less than 3 MeV for
these nuclei.

\section{Summary}
\label{sec:summary}

We presented an approximate and feasible approach to the TDHFB;
canonical-basis TDHFB method.
Since the number of the canonical states we need to calculate is
the same order as the particle number,
this method significantly reduces the computational task of the
TDHFB.
We calculated the $E1$ strength distribution in proton-rich Mg
isotopes, using the real-time real-space method.
We found a strong Landau damping effect in the drip-line nuclei,
that may be related to high level density of the background 1p1h
states with negative parity.
The calculation also indicates an increase of the low-energy $E1$ strength
as the nucleus approaches the proton drip line.

\ack
This work is supported by Grant-in-Aid for Scientific Research(B)
(No. 21340073) and on Innovative Areas (No. 20105003).
The numerical calculation was performed on RICC at RIKEN,
the PACS-CS at University of Tsukuba, and Hitachi SR11000 at KEK.

\section*{References}
\bibliographystyle{iopart-num}
\bibliography{nuclear_physics,myself}

\providecommand{\newblock}{}
\begin{thebibliography}{10}
\expandafter\ifx\csname url\endcsname\relax
  \def\url#1{{\tt #1}}\fi
\expandafter\ifx\csname urlprefix\endcsname\relax\def\urlprefix{URL }\fi
\providecommand{\eprint}[2][]{\url{#2}}

\bibitem{Neg82}
Negele J~W 1982 {\em Rev. Mod. Phys.\/} {\bf 54} 913--1015

\bibitem{NY05}
Nakatsukasa T and Yabana K 2005 {\em Phys. Rev. C\/} {\bf 71} 024301

\bibitem{Mar05}
Maruhn J~A, Reinhard P~G, Stevenson P~D, Stone J~R and Strayer M~R 2005 {\em
  Phys. Rev. C\/} {\bf 71} 064328

\bibitem{UO06}
Umar A~S and Oberacker V~E 2006 {\em Phys. Rev. C\/} {\bf 73} 054607

\bibitem{UO07}
Umar A~S and Oberacker V~E 2007 {\em Phys. Rev. C\/} {\bf 76} 014614

\bibitem{WL08}
Washiyama K and Lacroix D 2008 {\em Phys. Rev. C\/} {\bf 78} 024610

\bibitem{CC09}
Golabek C and Simenel C 2009 {\em Phys. Rev. Lett.\/} {\bf 103} 042701

\bibitem{BR86}
Blaizot J~P and Ripka G 1986 {\em Quantum Theory of Finite Systems\/}
  (Cambridge: MIT Press)

\bibitem{HN07}
Hashimoto Y and Nodeki K 2007
  arXiv:0707.3083

\bibitem{ASC08}
Avez B, Simenel C and Chomaz P 2008 {\em Phys. Rev. C\/} {\bf 78} 044318

\bibitem{Eba10}
Ebata S, Nakatsukasa T, Inakura T, Yoshida K, Hashimoto Y and Yabana K 2010
  arXiv:1007.0785

\bibitem{RS80}
Ring P and Schuck P 1980 {\em The Nuclear Many-Body Problems\/}
  (New York: Springer-Verlag)

\bibitem{BF76}
B{\l}ocki J and Flocard H 1976 {\em Nucl. Phys. A\/} {\bf 273} 45--60

\end{thebibliography}

\end{document}